\documentclass[10pt, conference]{IEEEtran}

\ifCLASSOPTIONcompsoc
  \usepackage[nocompress]{cite}
\else
  \usepackage{cite}
\fi
\usepackage{graphicx}
\usepackage{caption}
\usepackage{subcaption}

\ifCLASSINFOpdf
\else
\fi
\usepackage{lipsum}
\usepackage{tabularx}
\usepackage{url}

\begin{document}
%
\title{(Literally) above the clouds: virtualizing the network over multiple clouds}
%
%
%
%
%

%
\author{\IEEEauthorblockN{Max Alaluna, Fernando M. V. Ramos, Nuno Neves}
\IEEEauthorblockA{LaSIGE, Faculdade de Ciencias, Universidade de Lisboa, Portugal\\
malaluna@lasige.di.fc.ul.pt, fvramos@ciencias.ulisboa.pt, nuno@di.fc.ul.pt}
}

\maketitle
\begin{abstract}
Recent SDN-based solutions give cloud providers the opportunity to extend their ``as-a-service'' model with the offer of complete network virtualization.
They provide tenants with the freedom to specify the network topologies and addressing schemes of their choosing, while guaranteeing the required level of isolation among them.
These platforms, however, have been targeting the datacenter of a single cloud provider with full control over the infrastructure. 

This paper extends this concept further by supporting the creation of virtual networks that span across several datacenters, which may belong to distinct cloud providers, while including private facilities owned by the tenant.
In order to achieve this, we introduce a new network layer above the existing cloud hypervisors, affording the necessary level of control over the communications while hiding the heterogeneity of the clouds.
The benefits of this approach are various, such as enabling finer decisions on where to place the virtual machines (e.g., to fulfill legal requirements), avoiding single points of failure, and potentially decreasing costs.
Although our focus in the paper is on architecture design, we also present experimental results of a first prototype of the proposed solution.
\end{abstract}


\section{Introduction}



Virtualization gives the degree of flexibility necessary for cloud providers to achieve their operational goals while satisfying customer needs, by exposing a software abstraction to tenants (a Virtual Machine, VM) instead of the physical machine itself.
However, until recently, virtualization was restricted to compute and storage resources. Software Defined Networking (SDN) has proved to be a key enabling technology for network virtualization, as it can support logical communication endpoints coupled with on the fly data forwarding reconfiguration.
Newly proposed platforms~\cite{Koponen2014,AlShabibi2014,drutskoy2013} rely on SDN to offer full virtualization of the network topology and addressing schemes, while guaranteeing the required isolation among tenants.

These state-of-the-art platforms show the feasibility of network virtualization but they have been confined to a datacenter controlled by a single cloud operator.
This restriction can become an important barrier as more critical applications are moved to the cloud.
For instance, compliance with privacy legislation may demand certain customer data to remain local (either in an on-premise cluster or in a cloud facility located in a specific country).
This sort of requirement is particularly severe in the health and public administration (e.g., IRS) sectors, which normally need to resort to ad hoc approaches if they want to offload part of their infrastructure to the cloud. 
Being able to leverage from several cloud providers can potentiate important benefits.
First, a tenant can be made immune to any single datacenter or cloud availability zone outage by spreading its services across providers.
Despite the highly dependable infrastructures employed in cloud facilities, several recent incidents give evidence that they can still generate internet-scale single points of failures~\cite{csa-2013}.
Second, user costs can potentially be decreased by taking advantage of dynamic pricing plans from multiple cloud providers.
Amazon's EC2 spot pricing is an example, which was recently explored to  significantly reduce the costs on certain workloads when compared to traditional on-demand pricing~\cite{Zheng2015}.
As providers increase the support of dynamic prices, the opportunity for further savings increases with the user ability to move VMs to less costly locations.
Third, increased performance can also be attained by bringing services closer to clients or by migrating VMs that at a certain point in time need to closely cooperate.

In this paper, we propose a new architecture that allows network virtualization to extend across multiple cloud providers, including a tenant's own private facilities, therefore increasing the versatility of the network infrastructure.
In this setting, the tenant can specify the required network resources as usual but now they can be spread over the datacenters of several cloud operators.
This is achieved by creating a new network layer above the existing cloud hypervisor to hide the heterogeneity of the resources from the different providers while providing the level of control to setup the required (virtual) links among the VMs.
We follow an SDN approach, where the new network layer contains an Open vSwitch (OvS) that is configured by an SDN controller, in order to perform the necessary virtual-to-physical mappings and the set up of tunnels to allow the network to be virtualized.
Our preliminary experiments show that this extra level of indirection results in a relatively modest overhead in our target scenarios.
\section{Design requirements}
\label{requirements}

The network virtualization platform we propose leverages on network infrastructure from both public cloud providers and private infrastructures (or private clouds) of the tenants.
This heterogeneity impacts on the level of network visibility and control that may be achieved, affecting the type of configurations that can be pushed to the network, with obvious consequences on the kind of services and guarantees that can be assured by the solution.


On one extreme case, the public cloud provider gives very limited visibility and no (or extremely limited) network control, which is often the case with commercial cloud service providers (e.g., AWS).
Even in this case, these clouds offer a full logical mesh among local VM instances (i.e., they provide a ``big switch'' abstraction), which we can use to implement logical software-defined datapaths and thus present a virtual network to the tenant.
On the other extreme, full access may be attainable if the cloud is private (i.e., the datacenter belongs to the tenant).
This results in a flexible topology that may be (partially) SDN-enabled, where both software and hardware switching may be employed.

Considering this setting, we aim to fulfill three requirements in the design of our multi-cloud network hypervisor.
The first requirement is to have \emph{remote, flexible control over the network elements}.
Traditional networks' lack of such control has been identified as the main reason for the limitations of current forms of network virtualization~\cite{Koponen2014}. 


The second requirement is to offer \emph{full network virtualization}, including topology and addressing abstraction, and isolation between tenants.
For topology abstraction, different mappings should be created when the network is setup.
For instance, a virtual link can correspond to multiple network paths connecting the two endpoints.
In addition, tenants should have complete autonomy to manage their own address space of the virtual network.
Lastly, isolation between users should be enforced at different degrees.
A first level is attained by separating the virtual networks of the users and then hiding them from each other when they are deployed.
A second level is to prevent the actions of one user to influence the network behavior observed by the others.
For example, if one of the users attempts to clog a particular link, this should not cause a significant decrease on the bandwidth available to the other users.


Requirement number three is the ability to perform \emph{network snapshot and migration}.
After a virtual infrastructure is deployed and has been running for a while, the user might want to stop it and then restart it later on (e.g., to improve dependability or to minimize costs).
A fundamental service to achieve this goal is the ability to snapshot a VM at a particular instant.
In order to offer VM snapshot creation, our platform needs to capture the network state relevant to the VM and then update it after the restart.
In addition, our platform should have the ability to migrate a VM along with all network state associated.

\section{Architecture}
\label{preliminary_architecture}

\begin{figure}[h]
\centering
\includegraphics[width=.45\textwidth]{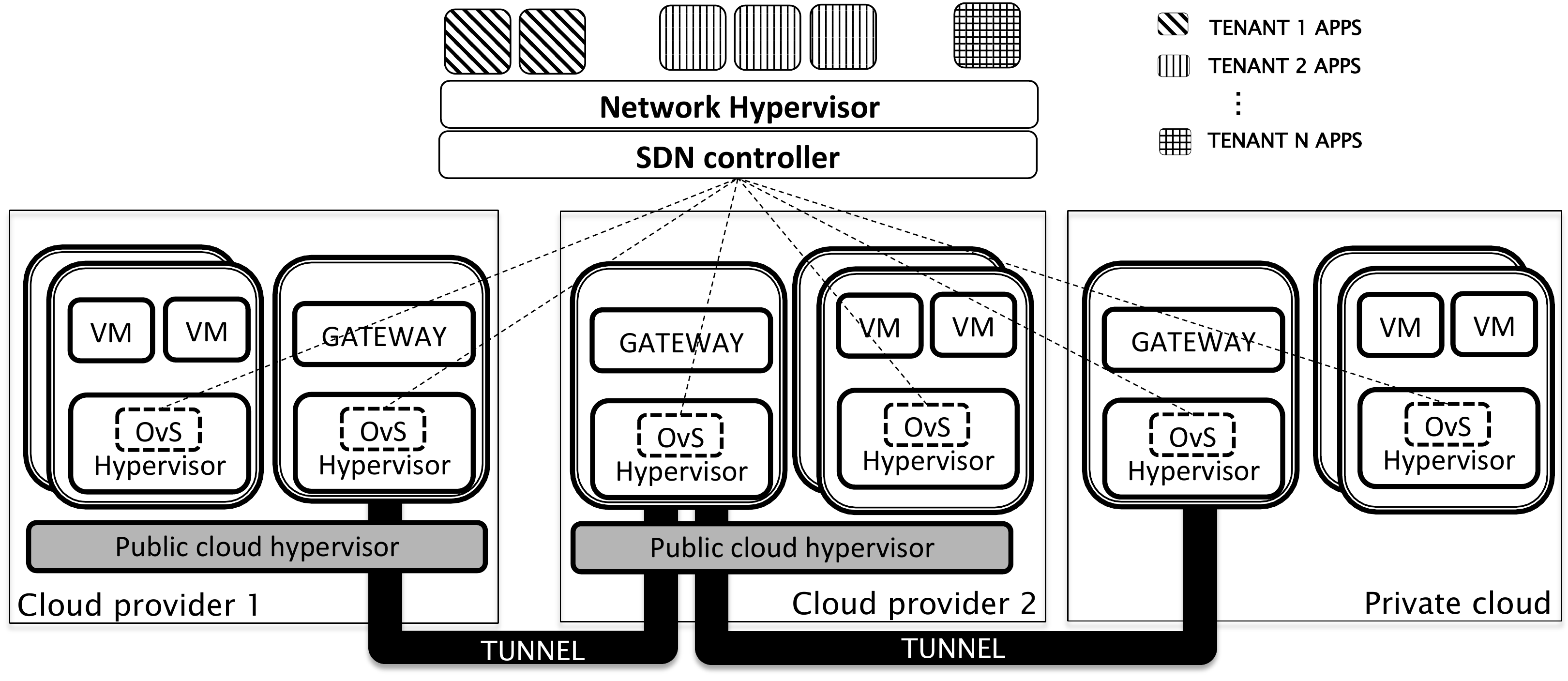}
\caption{Network virtualization architecture}
\label{fig:architecture}
\end{figure}

Recently proposed platforms for network virtualization~\cite{Koponen2014,AlShabibi2014,drutskoy2013} share a few characteristics.
First, they target datacenter environments where there is a high level of control over the resources.
Second, they rely on logically centralized control to achieve full network virtualization.
The novelty of our solution arises from tackling the challenges of using multiple clouds, including public clouds on which we have very limited control.   
The network virtualization architecture we propose is shown in Figure~\ref{fig:architecture}.

The network hypervisor controls and configures the OvS switches that are installed in all VMs (along with the OpenFlow hardware switches that may be present in the private cloud).
This hypervisor is built as an application that runs in the SDN controller, similar to NVP~\cite{Koponen2014} and FlowN~\cite{drutskoy2013}, and in contrast with the proxy-based approach followed by OpenVirteX~\cite{AlShabibi2014}.
Each cloud will have a specific VM, the gateway, that establishes tunnels with another clouds.
As such, only one public IP address per cloud is needed in our solution.
We build a minimum spanning tree to minimize the number of tunnels needed.
In a distributed configuration the gateway also hosts an instance of the SDN controller.
For each tenant, a specific set of network applications that control the tenants' virtual network runs on top of the network hypervisor.

The design of the architecture aims to fulfill the three requirements defined before.
To fulfill the first requirement -- network control -- the platform resorts to the SDN paradigm~\cite{Kreutz2015}.
The data plane element of our solution is OvS, a software switch for virtualized environments that resides within the hypervisor or management domain.
OvS exports an interface for fine-grained control of packet forwarding (via OpenFlow~\cite{mckeown2008}) and of switch configuration (via OVSDB~\cite{pfaff2013}).
This allows SDN-based logically centralized control.

In public clouds, our platform does not have access to the cloud hypervisor.
For this reason, we have an additional virtualization layer on top of the cloud hypervisor to provide virtualization between multiple tenants.
Our architecture supports both nested virtualization~\cite{Ben-Yehuda2010} and container-based~\cite{Vaquero2011} approaches for isolation between virtual machines. 
OvS is part of this hypervisor that runs inside each VM.
Private clouds include the proposed network hypervisor, with OvS, running on bare metal.

For the second requirement -- full network virtualization -- the network hypervisor has to guarantee isolation between tenants, while enabling them to use their desired addressing schemes and topologies. 
Our network hypervisor runs on top of the SDN controller to map the physical to virtual events by intercepting the flow of messages between the physical network and the users' applications.
This, along with flow rule redefinition at the edge of the network, allows isolation between tenants' networks.
For addressing virtualization, the traffic that originates from tenant VMs is all tagged.
The first 16 bits of the MAC address are used as tenant ID.
This design choice offers some advantages compared to using VLAN tags (the technique used by FlowN~\cite{drutskoy2013}, for instance), namely the fact that our solution represents a 10-fold increase in the number of tenants allowed.
In addition, our tenants can use the services provided by VLANs.
For topology abstraction we intercept all LLDP (Link Layer Discovery Protocol) messages.
LLDP is the protocol used in OpenFlow networks for network discovery.
By intercepting all topology-related messages the SDN controller can offer arbitrary virtual topologies to tenants.

Our platform also integrates VM snapshot and migration -- its third requirement.
Our system leverages on well-proven techniques for VM snapshotting and extends them for a multi-cloud setting.
In addition, we snapshot and migrate not only the VM, but also the network state.
The first step in snapshotting an SDN switch is the flow table, the list of pattern-action rules.
Flow table rules include not only the actions that match a certain packet header pattern, but may also include traffic counters and timers for deleting expired rules.
In addition to the flow table rules, the switch configuration and its queues are also snapshotted.

One option to migrate the tenant's VM and its virtual network could be to iteratively copy the VM and switch state, ``freeze'' the old network, and then start the new network.
This technique is undesirable, as freezing the network can lead to long outages.
To make live migration completely transparent to tenants we follow a cloning approach~\cite{Ghorbani2014}.
The idea is to clone one or more switches at a time, and then iteratively move the VMs associated, creating the necessary tunnels not to break connectivity.
This leads to two copies of the same switch to co-exist, potentially forwarding traffic and generating events at the same time.
To avoid inconsistencies it is necessary to limit, during the migration period, the switch actions taken autonomously (e.g., deleting rules after a timeout expires). 
In addition, to respect packet dependencies, specific rule updates need to be serialized.
This is done by setting the rule to temporarily send packets from this particular flow to the controller until it is guaranteed that the rules are installed in both switch replicas (e.g., using a barrier).   
Differently from~\cite{Ghorbani2014}, that targets a datacenter, our multi-cloud solution should perform efficiently in a (potentially) high-latency, bandwidth-constrained WAN environment. 

\section{Preliminary implementation and evaluation}
\label{evaluation}

The first prototype of our network hypervisor consists of nearly 4000 lines of Java code and is implemented as a module of the Floodlight controller.
GRE tunnels are used between the gateways, and a reactive SDN approach is used.
The flows rules are installed in the switches when the first packet-in is generated and sent to the controller. 

The evaluation of our solution answers two main questions.
First, it shows the cost of deploying the environment by analyzing the different components that make up the setup time.
In particular, we study how the creation of tunnels and the tunnel topology itself influence the setup time, and how this variable scales with network size.
Second, we evaluate the overhead introduced by our virtualization layer, both in the control and data planes.

The experiments were run on a testbed composed of two servers equipped with 2 Intel Xeon E5520 quad-core, HT, 2.27 GHz, and 32 GB RAM.
The hypervisor used is Xenserver 6.5, running OvS 2.1.3.
There is a router between the servers to simulate a multi-cloud environment.
One of the servers hosts one VM dedicated to the Floodlight controller, and another to host Mininet 2.2.0.

\emph{Setup time.} The setup time is the time between the moment the tenant submits a virtual network request until the instant when the whole network components are initialized and instantiated.
This time is composed of two components: time to populate network state in the resilient network hypervisor, and time to configure and initialize all tunnels.

\begin{figure}
    \centering
        \includegraphics[width=0.4\textwidth]{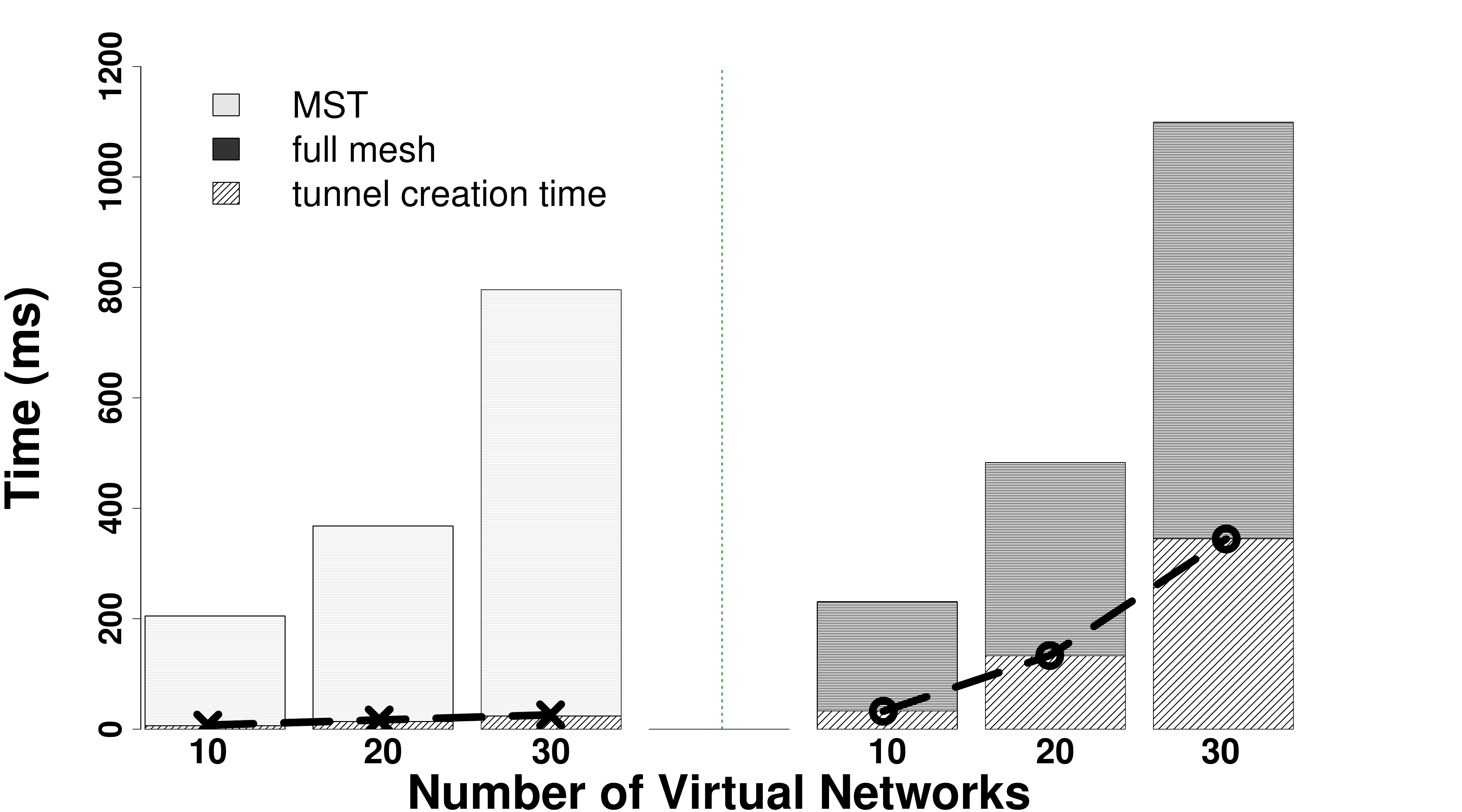}
    \caption{Setup time (left: MST; right: full mesh)}\label{fig:setup_time}
\end{figure}

We compare two different tunnel topologies.
The first is a setup with a full mesh of tunnels between all VMs, creating a one-hop tunnel between each pair of VMs, to serve as baseline.
The second is our solution: we set up a minimum spanning tree (MST) between those same VMs.
The results are shown in Figure~\ref{fig:setup_time}.

As expected, for the MST case the setup time grows linearly with network size.
By contrast, a full mesh has an O($n^{2}$) cost, and hence the setup time grows quadratically.
As can be seen in the full mesh case, tunnel creation has a visible effect on setup time as the network grows, making it a fundamental component for large scale scenarios.
This motivates the need to minimize the number of tunnels for the system to scale.
In any case, these setup times are still two to three orders of magnitude below the time to provision and boot a VM in the cloud~\cite{Li2010}.  

\begin{figure}
    \centering
        \includegraphics[width=0.4\textwidth]{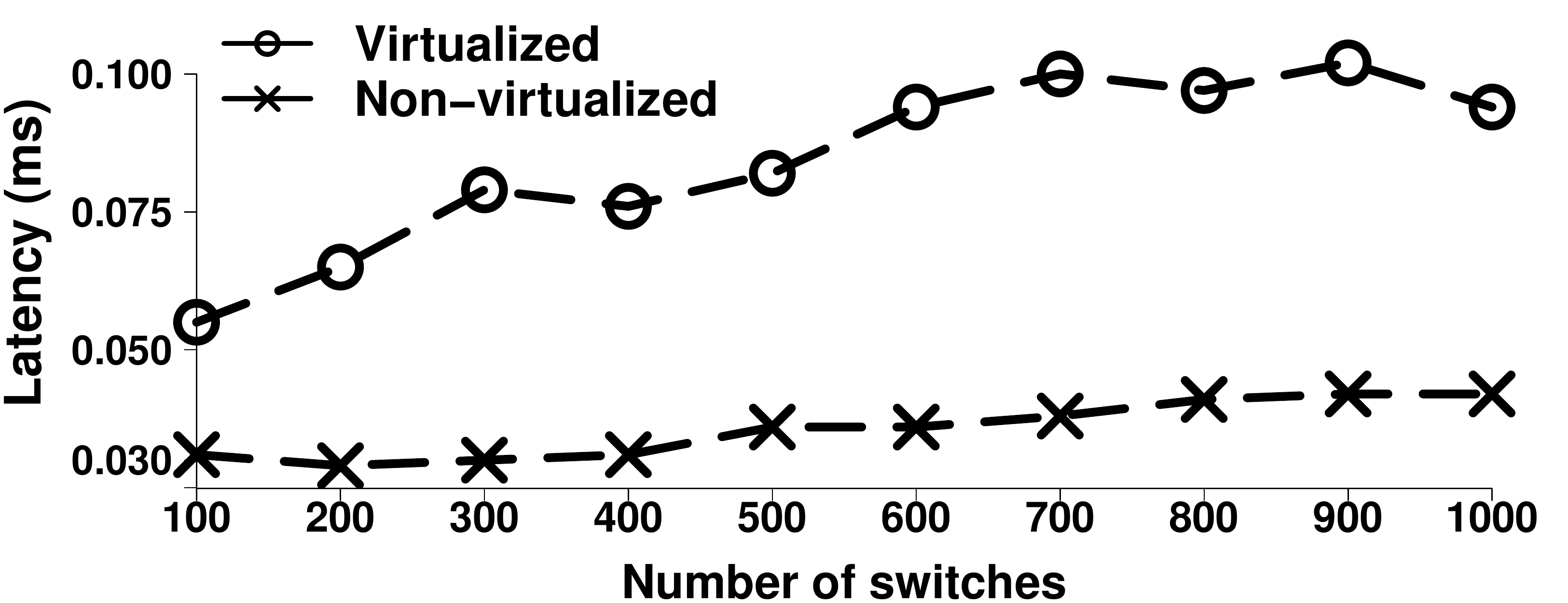}
    \caption{Control plane overhead}\label{fig:control_plane_overhead}
\end{figure}

\emph{Control plane overhead.} We measure the cost of network virtualization in the control plane using cbench, a control plane benchmarking tool that generates packet-in events for new flows.
In this test, cbench is configured to spawn a number of switches equal to the number of virtual networks, each switch having 5 hosts with unique MAC addresses.
The tests are run with cbench in latency mode.
In this mode cbench sends a packet-in request and waits for a response before sending the next request.
This allows measuring the controllerÕs request processing time.
We consider two scenarios: one with network virtualization, and another without network virtualization.
We present the results in Figure~\ref{fig:control_plane_overhead}.

As can be seen, the virtualization layer adds a very small overhead of less than 0.1 ms compared to the baseline.
Importantly, the latency overhead is mainly independent of network size (i.e., as the network grows the latency overhead remains relatively stable).
Further, for multi-cloud scenarios the inter-cloud latency is in the order of the tens of hundreds of milliseconds~\cite{Li2010}, and hence this overhead is negligible.

\begin{figure}
    \centering
        \begin{subfigure}[b]{0.37\textwidth}
        \includegraphics[width=\textwidth]{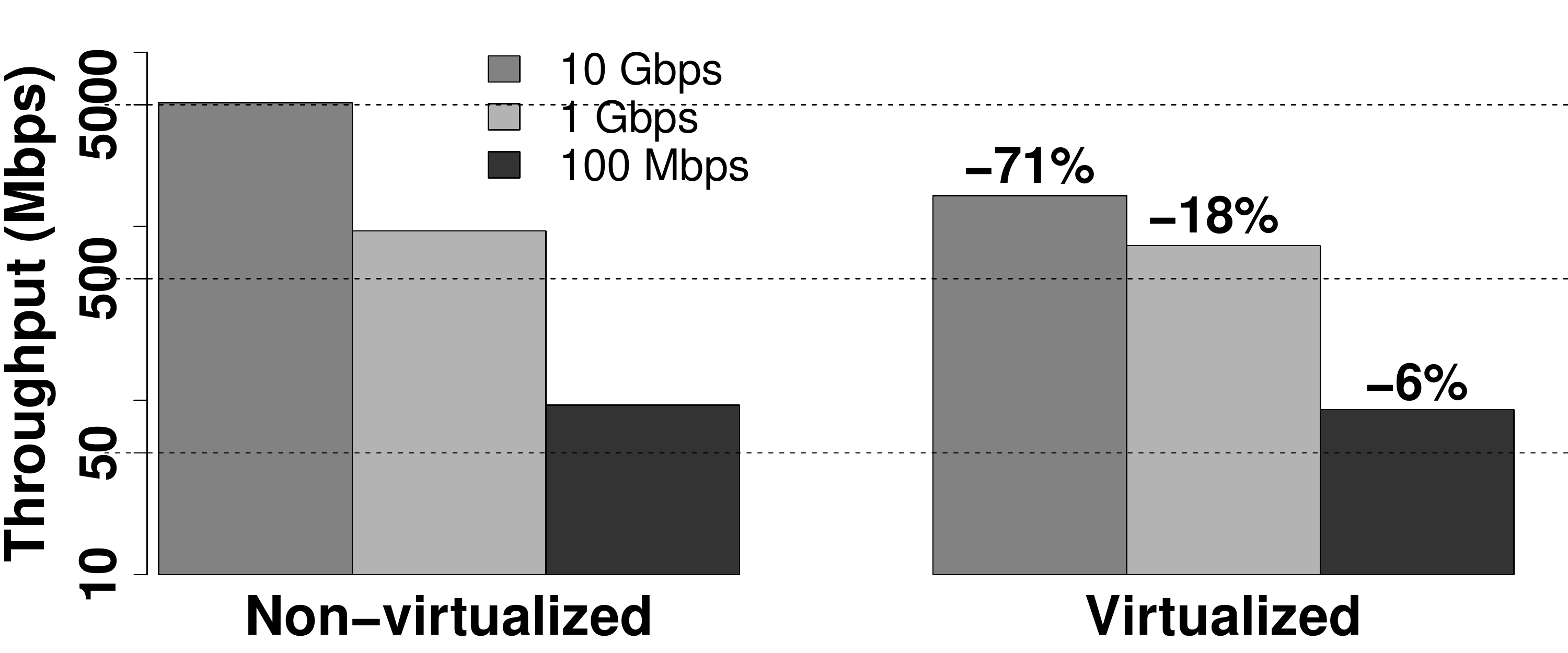}
        	\end{subfigure}
    \begin{subfigure}[b]{0.37\textwidth}
        \includegraphics[width=\textwidth]{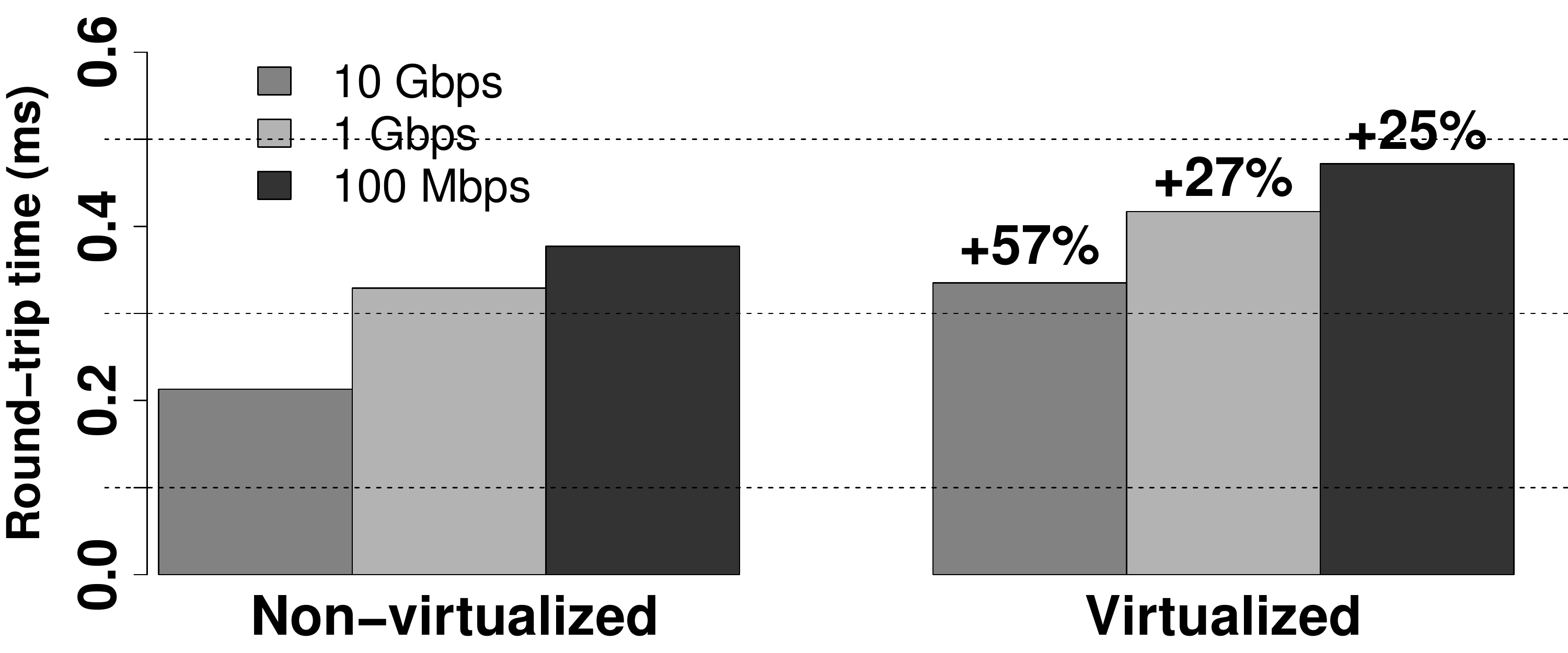}
    \end{subfigure} 
    \caption{Data plane overhead}\label{fig:data_plane_overhead}
\end{figure}

\emph{Data plane overhead.} To evaluate data plane overhead we make two experiments.
We measure network latency by running several pings between two virtual machines executing in different servers (emulating different clouds).
To measure network throughput we run netperf's TCP\_STREAM test between those same virtual machines.
Again, we consider two scenarios: one virtualized and one non-virtualized.
The results are shown in Figure~\ref{fig:data_plane_overhead}.

The results show that the virtualization layer introduces an overhead, in particular at very high bit rates.
This overhead is mainly due to the use of tunnels.
This motivates the need to minimize the use of tunnels by increasing traffic locality as much as possible.
This can be done by maintaining VMs that communicate frequently closer to each other.
For instance, VM migration could be triggered when this type of communication pattern is detected. 
Anyway, for the multi-cloud scenarios we target the inter-cloud throughput is in the order of the hundreds of Mbps~\cite{Li2010}.
At these rates, the overhead is relatively low.
The additional latency is also negligible when compared with typical inter-cloud latencies. 

\section{Conclusions}
\label{conclusions}

In this paper we have proposed the architecture of a network virtualization platform that spans across multiple cloud providers.
Such multi-cloud solution allows a tenant to be made immune to any single cloud outage, reduce costs by taking advantage of pricing plans from multiple cloud providers, and increase performance by bringing services closer to clients.
We introduce a new network layer above the existing cloud hypervisor to hide the heterogeneity of the different clouds.
SDN-based logically centralized control is used to offer full network virtualization. 
We have focused our discussion on the architecture, namely on its requirements and the techniques used to fulfill them.
In addition, we presented a first prototype, along with an evaluation aimed primarily at understanding the overhead introduced by the virtualization layer.
As work in progress, we are improving our platform to allow tenants to use the full header space (both L2 and L3) by rewriting IP and MAC addresses, instead of using a subset of the MAC address as tenant identifier.
We are also implementing and evaluating the techniques included in the architecture for network migration in a multi-cloud environment.

\section*{Acknowledgments}
This project has received funding from the European Union's Horizon 2020 research and innovation programme under grant agreement No H2020-643964 (SUPERCLOUD), and by national funds through Funda\c{c}\~ao para a Ci\^encia e a Tecnologia (FCT) with reference UID/CEC/00408/2013 (LaSIGE).

\bibliographystyle{unsrt}
\bibliography{references}  
%
%
\end{document}